\documentclass[]{aa}
\usepackage{graphicx}
\usepackage{amssymb}
\usepackage{natbib}
\bibpunct{(}{)}{;}{a}{}{,}

\begin{document}

\title{Bright CO ro-vibrational emission lines in the class I source GSS 30 IRS1}
\titlerunning{Bright CO ro-vibrational lines in GSS 30 IRS1}

\subtitle{Probing the inner disk of a young embedded star}
\author{K. M. Pontoppidan \inst{1} \and F. L. Sch{\"o}ier \inst{1} \and E. F. van Dishoeck \inst{1} \and E. Dartois \inst{2}}
\institute{Leiden Observatory, P.O.Box 9513, NL-2300 RA Leiden, The Netherlands \and Institut d'Astrophysique Spatiale, B{\^a}t. 121, Universit{\'e} Paris XI, 91405 Orsay Cedex, France}
\offprints{Klaus Pontoppidan,\email{pontoppi@strw.leidenuniv.nl}}
\date{Received / Accepted}

\abstract{We present a $\rm 4.5-4.85~\mu m$ $R=5\,000$ spectrum of the low mass class I young stellar object GSS 30 IRS1 ($L=\rm 25~L_{\odot}$)
in the $\rho$ Ophiuchus core, observed with the infrared spectrometer (ISAAC) on the {\it Very Large Telescope} (VLT-UT1).
Strong line emission from the ro-vibrational transitions of $\rm ^{12}CO$
and $\rm ^{13}CO$ is detected. In total more than 40 distinct
lines are seen in the covered region. The line emission is spatially extended
and detected up to $\rm 2\arcsec = 320~AU$ from the central source but is spectrally unresolved ($\Delta v < 30~\rm km~s^{-1}$).
This is the first time strong emission in the fundamental ro-vibrational band from CO has been observed from an embedded young stellar object.
The line fluxes were modeled using a 1-dimensional full radiative transfer code, which
shows that the emission is fully consistent with a gas in LTE at a single 
well constrained temperature ($T=515\pm5~\rm K$). Furthermore, the ratios
between lines from the two detected isotopic species of CO show that
the $\rm ^{12}CO$ lines must be optically thick. However, this is 
inconsistent with the observed spatial extent of the emission, since this
implies such low CO column densities that the lines are optically thin. A likely solution to the discrepancy is that the lines
are emitted by a smaller more dense region and then scattered 
in the bipolar cavity present around the central star. This gives a rough estimate
of the total molecular gas mass of $\rm 1-100~M_{\oplus}$ and a physical extent of 
$\rm\sim20-100~AU$. 
We propose that the most likely origin of the line emission is post-shocked gas in a dense dissociative accretion shock from the inner 
$\rm 10-50~AU$ of a circumstellar disk. The presence of a shock capable of dissociating molecules in the disk will have
implications for the chemical evolution of disks around young low mass stars.
\thanks{Based on observations obtained at the European Southern Observatory, Paranal, Chile, within the observing program 
164.I-0605. }
}

\maketitle

\section{Introduction}
The innermost $\rm 50~AU$ of the circumstellar environments around embedded young stellar objects is poorly constrained observationally due
to the large extinction through the embedding material ($A_V>20~\rm mag$) and the small angular size ($<1\arcsec$) of the region. The usual molecular
probes in the millimeter-submillimeter region are not effective for the temperatures ($T>200~\rm K$) and densities ($n\rm_{H_2}>10^7~cm^{-3}$) thought to be
present. However, an understanding of the processes taking place in this regime is essential to obtain a complete picture of the process of accretion
and the driving of outflows from low mass protostars as well as the early chemical and physical evolution of circumstellar disks.  

One of the most effective probes of warm dense gas is through emission in molecular ro-vibrational transitions. The most common bands readily available 
from ground-based facilities are the CO overtones and $\rm H_2$ fundamental bands around $\rm 2.2~\mu m$ \citep[e.g.][]{Reipurth} 
and the fundamental transitions of CO in the $M$-band around $\rm 4.7~\mu m$. Also emission from very hot water gas ($\sim 2000~\rm K$) near 
$2.29~\rm \mu m$ has been reported toward a few sources \citep[e.g.][]{Najita2}. Since the upper levels of these
transitions lie at temperatures of up to a few thousand Kelvin, they probe hot gas with temperatures between 100 and $\rm 1000~K$, making them
ideal to study the region of interaction between disk, protostar, accretion and outflow.
With the new generation of sensitive ground-based high resolution spectrometers for the $\rm 3-5~\mu m$ region an efficient window has been opened
for the detailed study of the CO ro-vibrational lines toward embedded young stellar objects (YSO). 
The past generation of instruments was suitable to either observe a low resolution spectrum with a 
fairly wide spectral range \citep[e.g.][]{teixeira} or a high resolution echelle spectrum with a very narrow spectral range \citep[e.g.][]{carr}. The main
exception are the Fourier Transform Spectroscopy (FTS) observations of \cite{mitchell,mitchell4}, who observed a number of high-mass stars in the 
entire M band at high spectral resolution ($R>10^6$), but such studies are limited to the brightest objects.
VLT-ISAAC has a large instantaneous spectral range in the $M$-band ($0.237~\mu m$), which combined with a medium resolution of 
$R=5\,000 - 10\,000$ and a limiting magnitude of $M\sim 10$ allows the entire fundamental band of gaseous CO of a 
wide range of young stars to be observed in a short time,
including low mass stars down to a few tenths of a solar mass in the nearest star-forming clouds. 

The embedded stars studied so far in CO ro-vibrational bands have showed mostly lines in absorption \citep{mitchell4,adwinElias29,adwinL1489} implying that
the warm CO gas is seen in front of a bright infrared continuum, produced by hot dust close to the central object. The fundamental CO lines are 
usually only seen in emission towards more evolved sources characterized by a class II type spectrum where
a circumstellar disk is directly visible \citep{carr,Blake}. CO overtone bandhead emission at $\rm2~\mu m$ has been observed 
in emission toward a few intermediate
mass pre-main sequence stars \citep{Thompson, Najita} and T Tauri stars \citep{carr1989}, and has been associated with hot gas 
($T\sim 1\,500-5\,000~\rm K$) located within a fraction of an AU in a Keplerian disk. 

We present here the peculiar $\rm 4.5-4.8~\mu m$ spectrum of the illuminating source IRS1=\object{Elias 21} of the bipolar 
reflection nebula \object{GSS 30} located in the
core of the $\rm \rho$ Ophiuchus molecular cloud at a distance of $\rm 160~pc$. It has been classified as a low mass class I YSO from
its spectral energy distribution (SED) \citep{GSS, Elias, WLY} and low bolometric luminosity \citep[$\rm L_{bol}=21-26~L_{\sun}$, ][]{GWAY,Bontemps}.  
Extensive polarimetric studies in the H and K band of the reflection nebula have shown that the source is surrounded by a large ($\rm\sim2\,000~AU$) 
disk-like envelope and a smaller circumstellar disk of $\rm\sim150~AU$, which are
inclined about $25\degr$ away from the plane of the sky, i.e. close to edge-on \citep{Chrysostomou,Chrysostomou2}. The high degree of linear polarization
(up to 50\%) as well as the presence of circular polarization imply that the light coming from the reflection nebula must have been multiply 
scattered, before heading into the line of sight. 

Two other sources (IRS2 and IRS3) are present toward the $K$-band reflection nebula (within 20\arcsec  of IRS1). 
IRS2 has a class III SED and is
probably a more evolved star \citep{Andre}. IRS3 has a class I SED, but is much fainter than IRS1 in the near-infrared. 
It has a bolometric luminosity of $0.13~L_{\odot}$, \citep{Bontemps}. IRS3 shows strong $\rm 6~cm$ emission 
\citep[see][where IRS3 is designated \object{LFAM 1}]{Leous}.  
   
The presence of a molecular outflow has not been firmly established. High velocity red- and blue-shifted CO millimeter emission to the south of GSS 30 has previously 
been reported by \citet{Tamura}, but since both lobes are located to the south of the infrared source, the gas is likely to
be associated with the \object{VLA 1623} jet, which is passing only $30\arcsec$ to the SW of IRS1. Using millimeter interferometric line data, \citet{Zhang} find  
evidence for a spherical expansion of the core surrounding the three sources in GSS 30. In addition, the inner region of the core seems to be cleared, which
is indicative of a young outflow. This is additionally supported by the presence of variable unbound water maser emission from within $0\farcs3$ of IRS1
\citep{Claussen}.

\section{Observations}
The observations of GSS 30 IRS1 were carried out using the long wavelength (LW) medium resolution mode on the Infrared Spectrometer
And Array Camera (ISAAC) mounted on the Very Large Telescope (VLT-UT1) at the Paranal Observatory on September
3, 2001. The detector for $\rm 3-5~\mu m$ observations is a $\rm 1K\times1K$ Aladdin InSb array, which allows a spectral coverage 
of $\rm 0.237~\mu m$ per setting in the $M$-band with a spectral resolution of up to $R=10\,000$. The detector resolution in the spatial direction
in the spectroscopic modes is $\rm 0\farcs148/pixel$. The spectrum of this source was taken as a part of a large program to study
ices and gas around young low mass stars \citep{VLTprogram}.

The data were obtained under excellent photometric conditions in service mode with $\sim$ 0\farcs3 infrared seeing and $\lesssim 10\%$ 
humidity which gave a very stable atmosphere. The frames were both chopped and nodded along the slit with a $15\arcsec$ chop throw.
The 0\farcs6 slit was used in two settings for 20 minutes per setting yielding a final signal to noise ratio on the 
continuum of the extracted spectrum ranging from 50 to more than 100 over a spectral range from $\rm 4.5~\mu m$ to $\rm 4.82~\mu m$ and with a 
spectral resolution of $R=5\,000$. 

The standard star \object{BS6084} (B1III) was observed at air masses between 1.2 and 1.5 immediately before GSS 30 for removal of telluric features, 
while the source was observed at air masses between 1.5 and 2.1. Because the source was already 
descending at the time of observation, it was not possible to observe source and standard at the same airmass as is otherwise of 
importance in order to obtain a good telluric subtraction. However, due to the exceptional stability of the atmosphere through most of the 
night we were able to correct for the airmass difference using a simple Lambert-Beer law, which states that the depth
of the atmospheric lines scales exponentially with the airmass. The final corrected spectrum turned out to
be of acceptable quality in spite of the airmass difference.
Our experience is that it is crucial to use a B star or an early A star as the standard because many narrow intrinsic stellar lines become visible in the 
$3-5~\mu$m region for spectral resolutions better
than $\sim 1000$ if an F or later type star is used.  

\begin{figure*}
\centering
\includegraphics[width=17cm]{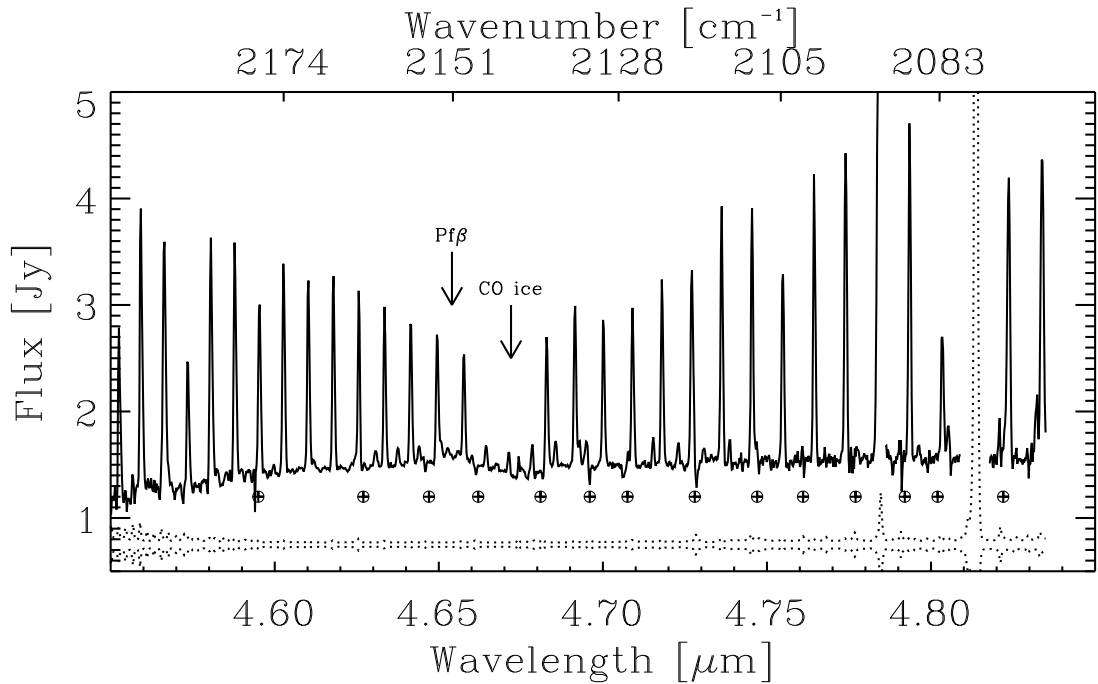}
\caption{The VLT-ISAAC $M$-band spectrum of GSS 30 IRS1 at $R=5\,000$. The statistical ($\rm 3\sigma$) error is shown below the spectrum. Note that this error
does not take the systematical errors introduced by variable telluric absorption into account which dominate in some parts of the spectrum. 
Since the peaks in the statistical error indicate deep atmospheric absorption they can be used to locate where telluric residuals
might be interfering with the intrinsic spectrum. The telluric residuals which are exceeding the noise are marked with a $\oplus$ and are seen to 
correlate well with the peaks in statistical error.}
\label{GSS 30spec}
\end{figure*}

The spectrum was reduced with IDL routines using standard methods appropriate for chopped and nodded infrared spectroscopy. 
Bad pixels were removed, the frames were distortion
corrected using the telluric features in a star-trace map as reference, and AB nodded pairs were subtracted to 
remove the long time scale differences which the chopping is not capable of removing. The spectra were extracted by integrating over a 
spatial width determined by the positions along the slit where the signal drops below the $3\sigma$ level. The raw spectrum was then ratioed by the standard 
after applying an optimized shift and airmass correction. In the final spectrum, the points where telluric residuals are apparent have been 
removed, including the region short-wards of $\rm 4.55~\mu m$, where atmospheric $\rm CO_2$ lines render the spectrum useless. The final result is shown in 
Fig. \ref{GSS 30spec}.

The CO ro-vibrational lines in the spectrum are red-shifted with $\rm \sim 20~km~s^{-1}$ 
due to the systemic velocity of the Earth relative to the source, which means that there is a significant
overlap between intrinsic CO lines and telluric absorption by CO. In spite of this, the telluric division is very good, especially when 
taking the unfavorable 
airmasses of source and standard into account. A few residuals from telluric $\rm ^{12}CO$ absorption are visible and are marked in the 
final spectrum. Since the resolution of our spectra is $\rm 60~km~s^{-1}$, none of 
the features in the source are unaffected by the telluric lines. At higher resolution ($R>20\,000$) the telluric lines may saturate, 
thereby forcing a division by zero when ratioing with the standard 
and resulting in a loss of information in the affected pixels. At sufficiently high resolution, this will not be a problem since the narrow
telluric lines will often be completely offset from the intrinsic lines.    However, at a resolution of 5\,000, the typical telluric line has a 
depth which is only a fraction of the intrinsic continuum allowing a significant signal to be present in the standard spectrum even in the 
center of a telluric line and only very few lines are partially or completely lost due to saturated telluric lines. 

By measuring the strength of the small residuals from telluric lines of species
not intrinsically present in the source spectrum, the systematic uncertainty in the source lines is estimated to be less than 15\%.
The statistical uncertainty for the brighest lines is typically a few \%, while the faintest lines are $3\sigma$ detections.
The spectrum was wavelength calibrated relative to the telluric lines and flux calibrated relative to the standard star. The wavelength
calibration of high resolution M band spectra is very accurate due to the large number of telluric lines, and a conservative estimate
for the precision is $\rm \sim 5~km~s^{-1}$ over the entire spectrum. Indeed, the standard deviation of the line center velocities given in Table 
\ref{COLineTable} is $\rm 4.2~km~s^{-1}$.  
Due to the difference in airmass between standard and source, we estimate the systematic uncertainty in absolute flux calibration to be $\sim 30$\%.
Since the bright lines are typically most severely affected by telluric residual, all line fluxes have total uncertainties with values 
between 15 and 50\%. The exact uncertainty for each line is difficult to estimate and we consequently adopt a value of 30\% for all lines.

\section{Results}

\subsection{Signatures of hot CO gas}
\label{HotGas}

It is evident that the $M$-band spectrum of GSS 30 is dominated by strong emission lines from the ro-vibrational ($v$=1--0) transitions of
$\rm ^{12}CO$ gas. R- and  P-branch lines within the observed spectral range are detected up to $J$=15 and $J$=18, respectively. The lines
are all unresolved at a resolution of $R=5\,000$, which gives an upper limit to the intrinsic line width of $\rm 30~km~s^{-1}$. No apparent 
decrease in line intensity is seen towards higher rotational quantum numbers, so observing the source in the entire $M$-band will clearly detect 
more lines. The brightest lines reach peak fluxes of more than 200\% of that of the continuum. However, since none of the lines are resolved, this is only a lower
limit to the peak flux. Fainter detected features include ${\rm ^{13}CO}~v$=1--0 lines as well as ${\rm^{12}CO}~v$=2--1 lines. The fact that the observed gas has 
a population excited to the second vibrational level already indicates that the gas is at least warmer than a few hundred K.     
The ${\rm ^{12}CO}~v$=1--0 lines have intensities of order $\rm 2\times10^{-13}~erg~s^{-1}~cm^{-2}$, the ${\rm ^{13}CO}~v$=1--0 
and ${\rm ^{12}CO}~v$=2--1 lines being 1-2 orders of magnitude fainter. The line fluxes of all unblended lines were extracted by fitting a gaussian as well 
as a first order polynomial to the local continuum. 
After subtraction of the continuum, the fluxes were derived by integrating the spectrum in a $3\sigma$ region around the line. 

The extracted fluxes of all unblended lines are given in Table \ref{COLineTable}. Many of the
faint lines are completely blended with the bright lines, but it was still possible to extract 12 ${\rm ^{13}CO}~v$=1--0 lines and 4 ${\rm ^{12}CO}~v$=2--1
lines, which is sufficient to constrain the most important parameters of the emitting gas. 

In addition to the CO emission lines, a faint 4.67 $\rm \mu m$ CO ice feature is visible. Since the feature is partly filled by the bright 
$\rm ^{12}CO~P(1)$ line it is not possible to obtain a well-defined ice profile. For the same reason the affected emission line is not
included in the analysis and modeling. There is also a very broad ($\rm FWHM\sim 500~km~s^{-1}$), $\rm 4.65~\mu m$ 
Pfund $\beta$ hydrogen recombination line visible, although with the bright CO lines superposed. 

Finally the $\rm H_2~0-0~S(9)$ line is detected with an integrated line flux of $\rm 2\pm1\times 10^{-14}~erg~s^{-1}~cm^{-2}$, signifying the
presence of hot ($\rm >1000~K$) molecular hydrogen or possibly shock-excited gas.

\begin{table}
\centering
\begin{flushleft}
\caption{Integrated CO line fluxes toward \object{GSS 30 IRS1}}
\begin{tabular}{lll}
\hline 
\hline 
Transition & Line flux & Heliocentric velocity\\
& [$\rm 10^{-13} erg~cm^{-2}~s^{-1}$] & [$\rm km~s^{-1}$]\\
\vspace{-0.35cm}\\
\hline 
\vspace{-0.35cm}\\
\multicolumn{3}{l}{${\rm ^{12}CO}~v$=1--0}  \\ 
\hline
$R(0)$ & 1.16 & -16\\
$R(1)$ & 1.27 & -13\\
$R(2)$ & 1.64 & -11\\
$R(3)$ & 1.74 & -11\\
$R(4)$ & 1.69 & -10\\
$R(5)$ & 2.08 & -11\\
$R(6)$ & 2.16 & -11\\
$R(7)$ & 2.49 & -10\\
$R(8)$ & 2.10 $\rm^a$ & -5\\
$R(9)$ & 2.78 & -10\\
$R(10)$ & 3.14 & -14\\
$R(11)$ & 1.46 $\rm^a$& -4\\
$R(12)$ & 3.85& -14\\
$R(13)$ & 3.42& -14\\
$P(2)$ & 1.59& -11\\
$P(3)$ & 1.58& -11\\
$P(4)$ & 1.62& -12\\
$P(5)$ & 1.73 $\rm^a$& -12\\
$P(6)$ & 1.94& -8 \\
$P(7)$ & 2.76 $\rm^a$&-3 \\
$P(8)$ & 2.84& -9 \\
$P(9)$ & 2.42& -10 \\
$P(10)$ & 2.16& -8 \\
$P(11)$ & 2.95& -2 \\
$P(12)$ & 3.37& -4 \\
$P(14)$ & 3.40& -7 \\
\hline
\vspace{-0.35cm}\\
\multicolumn{3}{l}{${\rm ^{12}CO}~v$=2--1} \\
\hline
$R(6)$ & 0.08& -17 \\
$R(7)$ & 0.10& -12 \\
$R(8)$ & 0.10& -14 \\
$R(9)$ & 0.07& 0 \\
\hline
\vspace{-0.35cm}\\
\multicolumn{3}{l}{${\rm ^{13}CO}~v$=1--0} \\
\hline
$R(3)$ &0.19& -11 \\
$R(4)$ &0.18& -11 \\
$R(5)$ &0.24& -14 \\
$R(6)$ &0.29& -11 \\
$R(10)$ &0.30& -15 \\
$R(11)$ &0.43& -14 \\
$R(12)$ &0.23& -14 \\
$R(13)$ &0.27& -10 \\
$R(16)$ &0.17 $\rm^a$& -15  \\
$R(17)$ &0.23& -3 \\
$R(18)$ &0.19& -19  \\
$R(21)$ &0.02& -15 \\
\hline
\end{tabular}
\label{COLineTable}
\end{flushleft}
\begin{list}{}{}
\item[$\rm^a$] This line is severely affected by telluric residual.
\end{list}
\end{table}

\subsection{Size of the emitting region}

\begin{figure}
\resizebox{\hsize}{!}{\includegraphics[width=8cm]{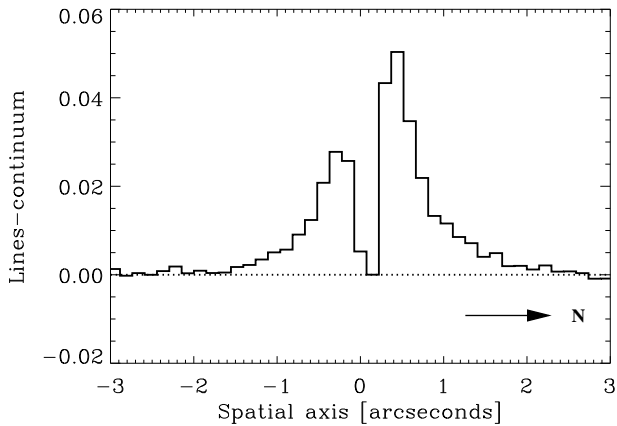}}
\caption{Spatial cross section of the resolved line emission. The 
spatial distribution of the emission (along the slit) is shown averaged over all ${\rm ^{12}CO}~v$=1--0 lines with the continuum spatial
profile subtracted. The line spatial profile was normalized to the continuum profile to show that the lines are detected
over a larger region than the continuum. The direction to the north on the sky is indicated on the figure.}
\label{COspatial}
\end{figure}

\begin{figure}
\resizebox{\hsize}{!}{\includegraphics[width=8cm]{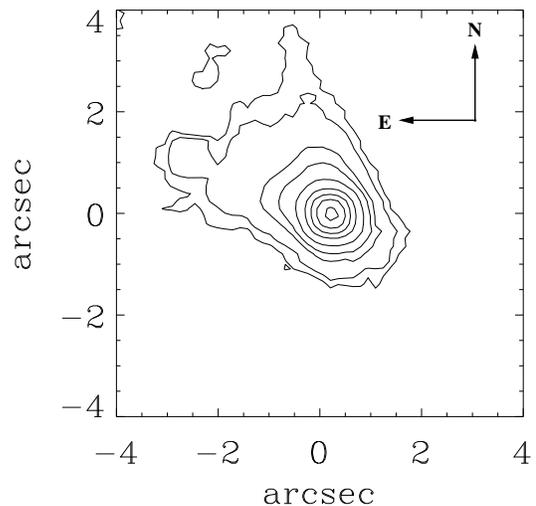}}
\caption{The $\rm 3.21~\mu m$ acquisition image of the reflection nebula surrounding GSS 30 IRS1. The north-eastern lobe is clearly visible and also 
extended emission from the south-western lobe is resolved. The slit was centered on $(0,0)$ and directed north-south. The contours are roughly logarithmic
with the lowest contour at values two orders of magnitude smaller than the highest contour.}
\label{GSS 30pic}
\end{figure}

In order to determine the parameters of the line emitting gas unambiguously, the size of the emitting region must 
be known. The FWHM of the continuum in the spatial direction is about $0\farcs5$, which is consistent with
the seeing throughout the observation. However, the lines appear to be spatially resolved, although most of the line flux is centered on the
source. The wings of the spatial
profile of each line extend up to $2\arcsec$ away from the central source along the slit. This is illustrated in Fig. \ref{COspatial}, 
which shows the normalized average of the spatial profile of the lines, with the normalized average of the continuum profile subtracted.  
It is evident that the line emission is resolved and detected to a distance of more than $\rm 320~AU$ from the central source. There is also an indication
that the northern part of line emission is stronger than the southern part, which agrees with the morphology seen in the $L$-band image of
IRS1 shown in Fig. \ref{GSS 30pic}. 

This extension suggests that the line emission is associated with the scattering nebula and is not coming directly from the central source. 
Since the continuum is less spatially extended than the line emission, the line emitting region appears 
to be spatially distinct from the continuum emitting region. Possible interpretations are that the line emission can be coming
from gas present in the bipolar lobes (e.g. an outflow), from resonance scattering of continuum emission or from scattering of lines emitted closer
to the star. Each of these possibilities will be discussed in detail below as the size of the emitting region depends strongly on the
scenario. 

\section{Models}
\subsection{Optically thin LTE models}

\begin{figure*}
\centering
\includegraphics[width=12cm,angle=90]{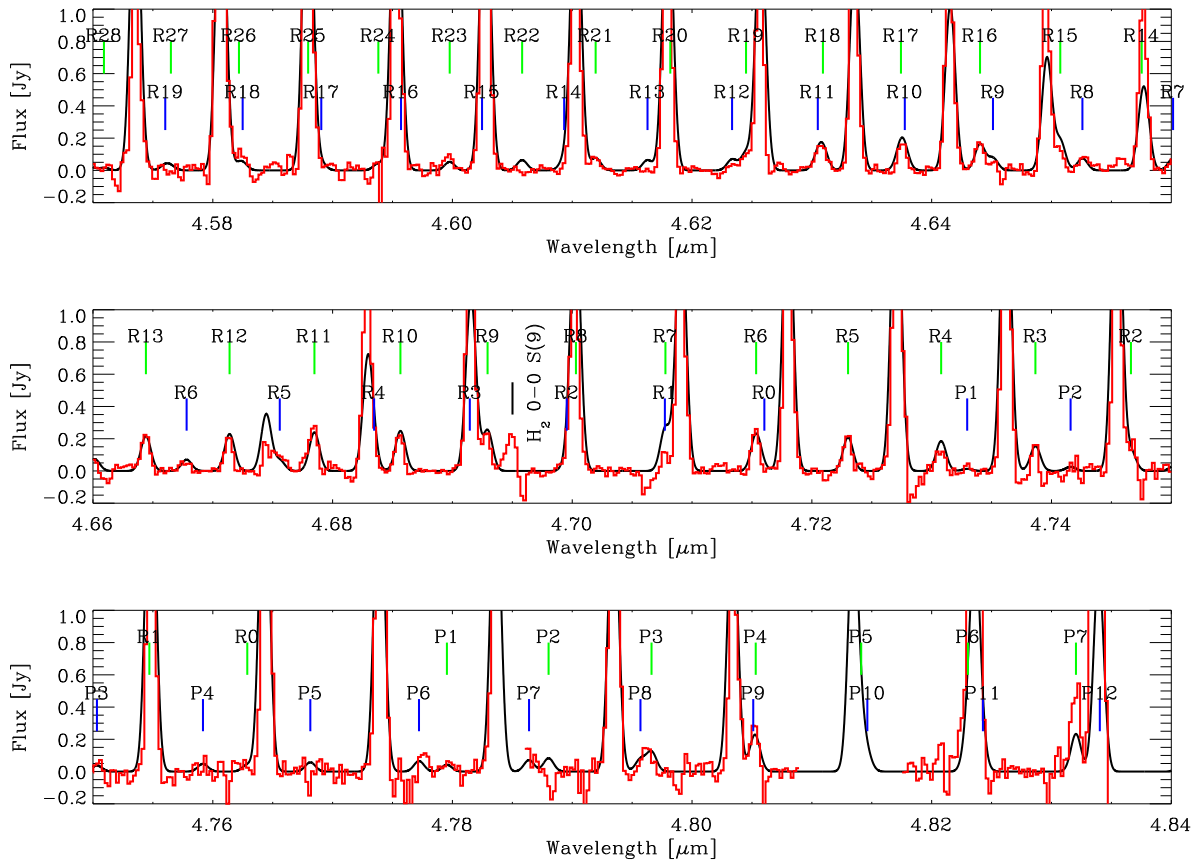}
\caption{Continuum subtracted spectrum of GSS 30 IRS1. Top row of line identifications show the ${\rm ^{13}CO}~v$=1--0 lines. Bottom
row of identifications show the ${\rm ^{12}CO}~v$=2--1 lines. The data have been over-plotted with the optically thin LTE model spectrum. 
The model ${\rm ^{12}CO}~v$=1--0 line fluxes have been divided by a factor of 7 for clarity and to illustrate the optical depth effects.}
\label{GSS30model}
\end{figure*}

As a simple first approximation, the extracted line intensities were fitted with a one-temperature optically thin LTE model. The
molecular data were taken from the HITRAN database \citep{HITRAN} and the relative line intensities were calculated directly from 
a Boltzmann level population distribution. The two parameters of the model, the gas kinetic temperature and the CO column density, should be well constrained
in the optically thin limit. If only the ${\rm ^{12}CO}~v$=1--0 transitions are used, an optically thin model is to some extent degenerate in the 
two parameters. This degeneracy is due to the fact that the populations in the vibrationally excited levels are much more sensitive to temperature changes
in the $\rm 400-800~K$ range than the rotational level populations. Since lines from two species are available, the degeneracy is broken.
It is reasonable to assume that the emission from both species is coming
from the same mass of gas, and that the $\rm ^{12}C/^{13}C$-ratio is close to 60 \citep[e.g.][]{Bensch}. 

This model is not able to provide a good fit to the entire line spectrum. 
However, from Fig. \ref{GSS30model} it is evident that the faint ${\rm ^{13}CO}~v$=1--0 and ${\rm ^{12}CO}~v$=2--1 lines can be simultaneously fitted
with a single temperature gas in the optically thin limit. The model lines of ${\rm ^{12}CO}~v$=1--0 are however an order 
of magnitude too bright. Excluding the ${\rm ^{12}CO}~v$=1--0 lines from the fit yields a well-constrained temperature of $\rm 515\pm25~K$ and a total
$\rm ^{12}CO$ column density of $\rm4\pm1\times10^{14}~cm^{-2}$. Since the fit includes lines from both CO isotopic species, we conclude that it is
unlikely that the apparent discrepancy between the ${\rm ^{12}CO}~v$=1--0 and ${\rm ^{13}CO}~v$=1--0 lines is due to a deviation
from the normal interstellar ratio. The simplest explanation is that the ${\rm ^{12}CO}~v$=1--0 lines are optically thick, 
while the rest remain optically thin. This would conveniently prevent us from seeing the whole
column of $\rm ^{12}CO$ and could possibly lower the line intensities enough to explain the data. However, the fundamental 
ro-vibrational transitions of $\rm ^{12}CO$ do not produce optically
thick lines for column densities less than $\rm \sim 10^{17}~cm^{-2}$, which is still 250 times more than the simple model dictates. 
Thus the optically thin LTE model is internally inconsistent.

\subsection{Results of the full radiative transfer model} 

\begin{figure*}
\includegraphics[width=18cm]{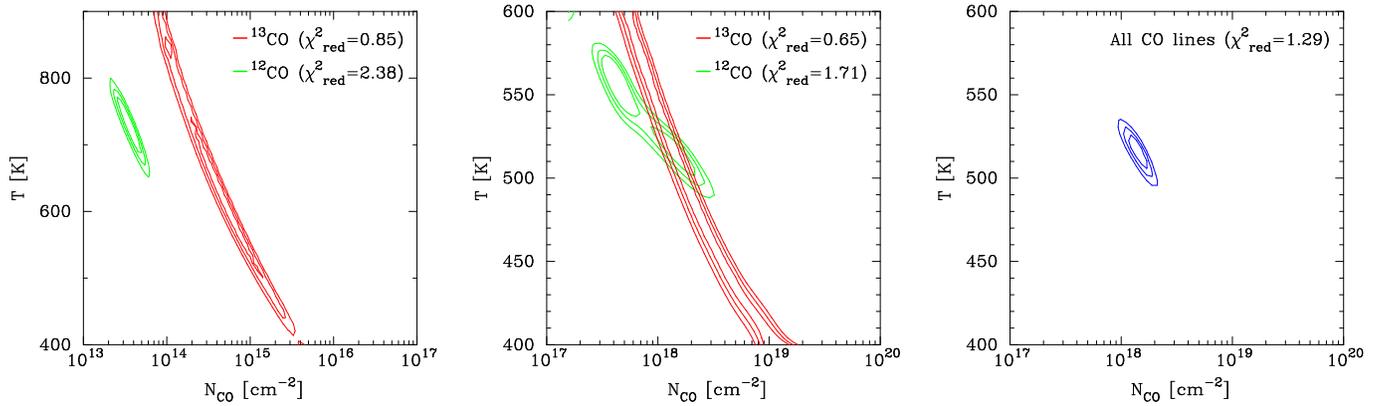}
\caption{$\chi^2$ maps of the best fitting models. Left panel: the ${\rm ^{12}CO}~v$=1--0 and ${\rm ^{12}CO}~v$=2--1 lines plotted against the ${\rm ^{13}CO}~v$=1--0 lines without
the assumption of scattering, in which case the sets of well-fitting parameters do not overlap. Middle panel: $\chi^2$ contours for the ${\rm ^{12}CO}~v$=1--0 and
${\rm ^{12}CO}~v$=2--1 lines over-plotted by contours for the ${\rm ^{13}CO}~v$=1--0 lines, but now with a scattering factor $\rm \epsilon=1.4\times10^3$. 
Right panel: The weighted sum of the two $\chi^2$ maps from the middle panel. The
contours show in all the maps the 1, 2 and 3$\sigma$ levels. }
\label{fredrikModel}
\end{figure*}

In order to quantify the discrepancy more accurately, we computed the full radiative transfer for a single temperature static envelope in LTE. 
For simplicity 
a spherical geometry was assumed since the results are not very sensitive to geometry. The size and density of the emitting region cannot 
be constrained by the model, so the quantities fitted were the single gas temperature and the column density of the gas, by keeping the size fixed at the
observed 4\arcsec.

Furthermore, it is
assumed that the line intensity is independent of the underlying continuum. This last assumption may not be completely valid, since 
the CO lines could be present in absorption in the continuum before the lines from the emitting gas are superposed. The effect
of this would be to lower the line peak flux with maximally the continuum flux level. Since the lines are completely unresolved, the true peak flux
of the lines is higher than that observed. The observed peak flux of the ${\rm ^{12}CO}~v$=1--0 lines is 2--3 times higher than the continuum
flux, so we conclude that this effect can only effectively lower the intensity of these lines with a small fraction at most. The fainter lines can be significantly 
suppressed by absorption in the continuum if they are optically thick, but this would only increase the discrepancy between the bright and faint
lines. Also, if the absorption lines are significantly broader than the superposed emission lines, the observed line intensities could be 
sufficiently suppressed. However, since we have no possibility of constraining this with the present data, we assume that the possible absorption lines 
and the emission lines have similar intrinsic widths, i.e. the contribution from absorption in the continuum is
negligible. Higher resolution spectra may clarify the matter.  

The ${\rm ^{13}CO}~v$=1--0 lines were first fitted independently of the ${\rm ^{12}CO}~v$=1--0 and ${\rm ^{12}CO}~v$=2--1 lines. The model confirms
that the two sets of lines cannot be fitted simultaneously, although each set is consistent with a single temperature gas. There is 
a certain degeneracy in the two free parameters, especially in the ${\rm ^{13}CO}$ lines, and the model gives a one-dimensional family of 
good fits for each species. However, the two 
families do not overlap for the observed line intensities. Fitting the ${\rm ^{12}CO}~v$=1--0 and the ${\rm ^{12}CO}~v$=2--1 lines simultaneously 
yields poorer fits than fitting them separately. All the lines are in the extreme optically thin limit with typical values of $\tau\simeq 10^{-3}$ for 
the ${\rm ^{12}CO}~v$=1--0
lines. The discrepancy between the ${\rm ^{12}CO}~v$=1--0 and ${\rm ^{12}CO}~v$=2--1 lines may be due to non-LTE effects, reflecting that the
rotational temperature $T_{\rm rot}$ and the vibrational temperature $T_{\rm vib}$ are not equal. However, non-LTE effects cannot explain why the 
${\rm ^{12}CO}~v$=1--0 to ${\rm ^{13}CO}~v$=1--0 line ratios differ with almost an order of magnitude from the predictions of the simplest LTE model, since the
excitation structures of different isotopic species must be almost identical. 

To explain the data, we propose that all the observed line emission has been scattered by dust in the bipolar cavity seen in the near-infrared. The
scattering efficiency is very low at $\rm 4.7~\mu m$, and thus only a small fraction of the original line emission will be 
scattered into the line of sight. 
This means that the line intensities from the emitting region and consequently the column densities can reach much larger values than those previously suggested 
by the model. To test this hypothesis, we adopted a factor, $\epsilon = S \times (\pi R^2)/\Omega$, which accounts for 
our lack of information about the true solid angle of the emitting region $\rm \Omega$, and the ratio of intrinsic line flux to line flux scattered 
into the line of sight, $S$. $R$ is the observed angular radius of the emitting region, which is taken to be $R=2\arcsec$. 
Absolute upper and lower bounds to the size of the emitting region are given by the natural conditions $S>1$ and $\Omega<\pi R^2$. 
Decreasing the size of the emitting region will increase the column densities, but since line emission is observed far from
the source, we cannot explain the entire discrepancy with a smaller emitting size. To recover the original flux it is necessary to 
multiply the observed values with this factor. Multiplying all observed line intensities with a 
sufficiently large factor will allow the ${\rm ^{12}CO}~v$=1--0 lines to become optically thick. We are able to constrain the scattering factor in the model
by demanding a simultaneous fit to all the lines. 

The best fitting value of the scattering factor is $\rm\epsilon=1.4\times 10^3$ with a reduced $\rm \chi^2$ of 1.29, which gives a well-constrained 
single temperature LTE gas
at $\rm 515\pm 5~K$ and $\rm N(CO)=1.5\pm 0.5\times 10^{18}~cm^{-2}$, as indicated in Table \ref{modelTable}. With an extent of the
emitting region of $4\arcsec$ an upper limit to the total gas mass of $\rm 200~M_{\oplus}$ is found. Since the total gas mass scales 
with $\rm \Omega^{0.5}$ an absolute lower limit to the mass is $\rm 1.3~M_{\oplus}$ with a diameter of the emitting region of
$\rm 0.19\arcsec = 30~AU$, assuming that $S=1$ and ignoring that line emission is observed to much larger distances from the central source. 
Since $S$ must be significantly less than unity, the total mass is most likely significantly larger than the lower limit.
It is not necessary to invoke any non-LTE effects to obtain a nearly perfect fit to all the observed lines and the assumption of
LTE holds such that $T_{\rm rot}=T_{\rm vib}$. The $\chi^2$ maps of the best-fitting models are shown in Fig. \ref{fredrikModel}.

It is possible to make an estimate for an absolute lower limit to the grain size by assuming $S=1$. In this case the scattering efficiency, 
$Q_{\rm sca} > \epsilon$, since it is not possible to scatter more light into the line of sight than out of it. This happens for silicate grain sizes larger than 
$\rm 0.2-0.3~\mu m$ \citep{DL,D}. As a consequence the grains must be large, perhaps as large as $\rm 1~\mu m$, which is expected
for grain growth in dense media \citep{Pendleton}. On the other hand, the lower limit on the size of the scattering grains does not 
fit with the polarization maps of the GSS 30 nebula. \cite{Chrysostomou} finds that grains larger than $\rm 0.35~\mu m$ are inconsistent
with the variation of linear polarization with wavelength. However, the polarization was mapped on a much larger angular scale than the 
size of the nebula at $\rm 4.6~\mu m$, so the discrepancy may be explained by grain growth within a few hundred AU of the central source.

\begin{table*}
\centering
\caption{Best fitting models for GSS 30 IRS1}
\begin{tabular}{lllll}
\hline
\hline
Model & Lines used & T & N(CO) & Red. $\rm\chi^2$  \\
& & [K] & [$\rm cm^{-2}$] \\
\hline
Optically thin direct calculation & ${\rm ^{13}CO}~v$=1--0 + ${\rm ^{12}CO}~v$=2--1 & $500\pm 25$ & $\rm 4\pm1\times 10^{14}$ & 1.4 \\
Full radiative transfer & ${\rm ^{13}CO}~v$=1--0 & degenerate & degenerate & 0.85\\
Full radiative transfer & ${\rm ^{12}CO}~v$=1--0 + ${\rm ^{12}CO}~v$=2--1& $725\pm50$ & $\rm 2.5\pm1\times10^{13}$ & 2.38\\
Full radiative transfer with scattering & All & $515\pm10$ & $\rm 1.5\pm0.5\times10^{18}$ & 1.29 \\
\hline
\end{tabular}
\label{modelTable}
\end{table*}

\subsection{Other scenarios and pumping mechanisms}

One way of avoiding the scattering assumption entirely may be to apply a two-temperature model to the data. In this scenario a small part of
the gas close to the central source with a kinetic temperature of $\rm \sim 1000~K$ would be responsible for the ${\rm ^{12}CO}~v$=2--1 lines 
as well as part of the high $J$ lines of the $v$=1--0 lines, while a massive outer component at $\rm \sim 200~K$ would dominate the emission in the low $J$
lines. The scattering assumption would not be needed since the ${\rm ^{12}CO}~v$=1--0 lines require a much higher column density to create the
observed line intensity at $\rm 200~K$ compared to $\rm 500~K$ and will thus more readily become optically thick. We applied a two-temperature model by
requiring that the total column density must be the same as that determined by the single-temperature model. This is reasonable because the optical 
depth of the lines is given by the ratio between the $\rm ^{12}CO$ and $\rm ^{13}CO$ lines independently of $\epsilon$
and therefore must be conserved. As in the single-temperature model three parameters are varied, namely the two temperatures and the ratio of 
the column density between the components. In other words, $\epsilon$ is exchanged as a free parameter with a second temperature, 
thus keeping the number of free parameters constant.

The $\rm ^{13}CO$ lines can be fitted well within the two-temperature model, although the parameters are somewhat degenerate in the
same way as in the single temperature model.
However, it was not possible to fit the $\rm ^{12}CO$ lines within this scenario, with the best reduced $\rm \chi^2$ being of order 10. 
Furthermore there is a preference towards a single low temperature in the $\rm ^{12}CO$ lines which is inconsistent with the
$\rm ^{13}CO$ lines. Inclusion of all lines gives no meaningful fits within the parameter space. 
The explanation is that it is not possible to combine the ${\rm ^{12}CO}~v$=2--1 and the ${\rm ^{12}CO}~v$=1--0 lines, since the 
high temperature component easily produces too much $v$=1--0 emission in order to fit the $v$=2--1 emission. 
We conclude that a two-temperature model is clearly inferior to the scattering model in explaining the observed line
emission.

Radiative pumping by a strong infrared continuum is another possibility that needs to be considered as an alternative excitation mechanism and a way
to avoid the scattering scenario by introducing non-LTE effects. This would produce a single vibrational temperature 
\citep{Scoville}. However it does not explain why the rotational and vibrational temperatures
seem to be identical. Also, luminosities larger than $\rm 10^3~L_{\odot}$ are required to effectively pump CO molecules to the second vibrational level
or higher at distances from the
central source beyond $\rm 1~AU$ \citep{Scoville}. Using the 1-dimensional Monte-Carlo radiative transfer code from \cite{FredrikThesis} illustrates the
effects of radiative excitation on the line populations. The simplest possible model for radiative excitation assumes a blackbody 
with a luminosity $L=25 \rm ~L_{\odot}$ at 
the center of a single density spherical and static envelope . The blackbody temperature is taken to be
the colour temperature of the continuum around $\rm 4.7~\mu m$, 
determined from the $M$-band spectrum to 
$\rm 420~K$. The radius used is the observed $\rm 320~AU$, the gas kinetic temperature is assumed to be a power law beginning at the
surface of the central source. The molecular constants are taken from \cite{FredrikConst}. This leaves the gas density as the only free parameter, 
which basically determines the
optical depth of the lines and consequently the importance of line trapping effects. The model calculations show
that a density of at least $n_{\rm H_2}\sim 10^6 ~ \rm cm^{-3}$ is needed to reproduce the approximate flux level of the ${\rm ^{12}CO}~v$=1--0 lines. The
vibrational excitation temperature is largely determined by the continuum flux level at $\rm 4.7~\mu m$ and therefore by the luminosity and to a 
smaller extent the temperature of the radiating source. Since the dominating mechanism for populating the $v$=2 level is via the $v$=1 level, and since
both the $v$=2--1 and $v$=1--0 transitions are placed in the same narrow wavelength region, the
radiative excitation rates are strongly dependent on local photon density in the $\rm 4.7~\mu m$ region. Consequently, as long as the
direct route through the $v$=2--0 overtone transitions around $\rm 2.35~\mu m$ is not important, the vibrational level population depends
on the temperature of the input radiation field only through the flux level at $\rm 4.7~\mu m$. 
At the adopted luminosity, the ${\rm ^{12}CO}~v$=2--1 line fluxes are two orders of magnitude too low, reflecting an excitation temperature
of $\rm 400~K$. To significantly populate the second vibrational levels, a luminosity of at least $\rm 10^3~L_{\odot}$ is required, in agreement with
\cite{Scoville}. Finally, the population of the rotational levels shows significant departures from a Boltzmann distribution, especially 
when line trapping becomes important. Altogether, this indicates that radiative pumping as an excitation mechanism in the case of GSS 30 IRS1 is unlikely.

Finally, the possibility that the lines are continuum emission which is resonance scattered in CO transitions should be considered. 
It is known that this takes place in the case of asymptotic giant branch stars, e.g. in the case of Mira \citep{Ryde}. However, since this mechanism
probes the scattering gas rather than the emitting medium, it will not solve the discrepancy. Also it would require that the gas at a distance
of $\rm 320~AU$ from the central source has a single temperature of $\rm 500~K$, which seems highly unlikely.

\subsection{Hydrogen recombination lines and mass loss rate}
\label{massloss}

\begin{table*}
\centering
\caption{Hydrogen recombination lines observed in GSS 30 IRS1}
\begin{tabular}{llll}
\hline
\hline
Transition & Line intensity & FWHM & Mass loss rate \\
& [$\rm erg~cm^{-2}~s^{-1}$] & [$\rm km~s^{-1}$] & [$\rm M_{\odot}~year^{-1}$]\\
\hline
$\rm Pf\beta$ & $1.3 \pm 0.5 \times 10^{-13}$ & $450 \pm 50$ & $1.7\times10^{-6}$\\
$\rm Br\alpha$ & $1.6 \pm 0.2 \times 10^{-13}$ & $\lesssim 500$ & $0.8\times10^{-6}$ \\
$\rm Pf\delta$ & $6.0 \pm 1.0 \times 10^{-14}$  & $\lesssim 500$ & $1.8\times10^{-6}$ \\
\hline
\end{tabular}
\label{HydrogenTable}
\end{table*}

There are a number of hydrogen recombination
lines common to YSOs ($\rm Pf\beta$, $\rm Br\alpha$, $\rm Pf\delta$) present in the $\rm 3-5~\mu m$ region, which are usually taken as an indicator of 
accretion activity. This is particularly true in the case of low mass young stars, where the hard radiation from the boundary layer between disk and star
is needed to produce the ionizing radiation, since the central star is not hot enough. Because these lines have broad wings ($\rm 3-500~km~s^{-1}$), they are
often interpreted as coming from a strong stellar wind, which may be the engine of an outflow. The possibility that the CO gas could be associated 
with a wind or molecular outflow should therefore be explored. If associated with a stellar wind, the lines could be formed either as 
cooling emission from the gas as molecules are reformed or as emission from a shock as the wind collides 
with the circumstellar envelope or disk.
It is known that no large scale outflow is present which could be produced by any of the three stars in the cluster, but there is some evidence 
of an expanding core surrounding the cluster from interferometric observations of rotational $\rm ^{13}CO$ and $\rm C^{18}O$ lines \citep{Zhang}. 

Hydrogen recombination lines emitted from a plasma moving at supersonic speeds can be modeled in the Sobolev approximation (Sobolev 1960), which is valid 
when the thermal line broadening is much smaller than the wind speed. 
This formalism has been treated for spherical ionized winds in LTE from evolved stars in \cite{Castor} and \cite{KS},
and for Herbig AeBe stars in \citet{Nisini}. Using this formalism an ionized mass loss rate can be estimated for
the star if a wind origin for the hydrogen lines is assumed. The extinction corrected line intensities for the hydrogen recombination lines 
observed in GSS 30 IRS1 are tabulated in 
Table \ref{HydrogenTable} along with
the mass loss rates derived from each line. The line ratios and the mass loss rate are not sensitive to the temperature, so a constant electron temperature
of $\rm 10^{4}~K$ has been assumed. Finally, it is assumed that the central star does not shadow a significant part of the radiation from the wind.  
Changing the model parameters, i.e. the temperature and the size of the ionized region, with a factor of two changes
the mass loss rate with a factor of two at most. The velocity structure is such that the wind starts at the surface of the star with $\rm 20~km~s^{-1}$
and rises quickly to $\rm 450~km~s^{-1}$ to fit the observed FWHM of the $\rm Pf\beta$ line. The exact shape of the velocity structure is otherwise not
significant for the derived mass loss rate. 

We find that an ionized mass loss rate of $\rm 1.7 \times 10^{-6}~M_{\odot}~year^{-1} = 0.6~M_{\oplus}~year^{-1}$ is consistent with the two 
observed Pfund lines. The $\rm Br\alpha$ line is too faint for this mass loss rate, which may indicate deviations from LTE. We can rule 
out that it is an effect from a poor extinction correction
since the  $\rm Br\alpha$ line lies between the $\rm Pf\beta$ and $\rm Pf\delta$ lines in wavelength. The most serious problem with this model
is likely that very little is known about the structure of winds from low mass embedded YSOs; in particular, the ionization structure will depend
directly on energetic processes which are poorly understood, such as accretion activity or interactions of the circumstellar matter with strong
magnetic fields. Also, if the wind is predominantly neutral, then the total mass loss rate will be much higher than the ionized mass loss rate and
will become inconsistent with the non-detection of an outflow. It is possible that other mechanisms such as accretion flows must be considered for
the formation of recombination lines from low mass young stars. 
The hydrogen lines are therefore not likely to provide any strong constraints on the emitting CO gas and a detailed discussion on the hydrogen emission
is outside the scope of this paper.

\section{Discussion}

In summary, to determine the origin of the CO gas emission it must be explained why only a single well-defined temperature is needed.
Which mechanism can heat up to $\rm 100~M_{\oplus}$ of gas to a unique temperature 
of $\rm 500~K$, yet
keep the intrinsic line width less than $\rm 30 ~km~s^{-1}$? Finally, why is no other similar embedded source from the literature 
showing the same strong CO rovibrational emission as GSS 30 IRS1?

\subsection{Outflow?}

If the warm CO gas resides in the wind/outflow component of the circumstellar environment 
then an estimate of the time needed to create it can be found
from the mass loss rate derived in Section \ref{massloss} under the assumption that the wind is predominantly ionized. 
A mass between 3 and 100$~M_{\oplus}$  corresponds to a production time of minimally 5 years 
and maximally a few centuries, while
the most likely time is about one century. It seems unlikely that hot, ionized material emitted over a longer period should thermalize at a 
single temperature. Also, an outflow origin of the line emission would produce broad wings in the lines, which is not observed.
If the wind is predominantly neutral, the resulting mass loss rate would be so high that a clear outflow should have been detected.

\subsection{Inner disk?}

Another possibility is that the emission is produced by warm thermalized gas in the disk itself.
Since more evolved T Tauri disks are known to exhibit similar CO emission, although with smaller intensities, it is conceivable that 
we are seeing an equivalent process in the case of a younger disk. 
A typical circumstellar disk around a low mass young star is often observed to have a mass of a few times the minimum solar nebula, i.e. 
$\rm M_{disk} \sim a~few~ 0.01~M_{\odot}$ \citep{Andre,Osterloh}. The disk mass for GSS 30 IRS1 inferred by the 1.3 mm continuum emission
is $M_{\rm disk} = 0.03~M_{\odot}$ \citep{Andre}. In hydrostatic equilibrium, a disk irradiated by the central star has a surface density $\Sigma\sim R^{-3/2}$
\citep{CG}. If viscous dissipation in the disk is taken into account the surface density attains a flatter $R$-dependency, $\Sigma\sim R^{-1}$. In both
cases the accumulated disk mass reaches the observed $\rm 10-100~M_{\oplus}$ between 2 and 10~AU from the central star. A single temperature of $\rm 515~K$
is however not consistent with the disk models, which prescribe a large range in temperatures with a small 1000-2000~K component within a few stellar
radii to an extended 100-200 K component within a few AU of the central star. However, in no models for passive circumstellar disks is it expected that
the disk exhibits a single temperature, indeed quite the contrary is the case. Furthermore, Keplerian rotation within a few AU would produce observable line
broadening. The narrow lines could only be explained if we are seeing light emitted vertically from the plane of the disk before being scattered
into the line of sight, effectively removing the effects of rotation. 

\subsection{Accretion shock?}

A final option is that the emission lines are cooling lines from post-shocked dense gas. Vibrational $\rm H_2$ emission is one of the principal tracers of
low density ($n\rm_{H_2}\lesssim 10^7~cm^{-3}$) shocked gas. As mentioned in Section \ref{HotGas} the 
(0--0) $\rm S(9)$ line of $\rm H_2$ at $\rm 4.695~\mu m$ is seen in our $M$-band spectrum, while no $\rm H_2$ lines are seen in the
$K$-band spectrum of GSS 30 IRS1 by \cite{GL} although no upper limit is given. Assuming that the observed $\rm H_2$ (0--0) $\rm S(9)$ line is thermalized at the 
CO temperature, the required molecular gas mass is $5~ \rm M_{\oplus}$. This is assuming that 
the $\rm H_2$ emission is seen directly and is not corrected for extinction, since the extinction is hard to estimate for embedded stars.
A typical Ophiuchus extinction of $A_{\rm V}=25~\rm mag$ \citep{teixeira2} will increase the molecular gas mass by a factor of 2. 
If the emission in the $\rm H_2$ line is scattered in the same way as the CO emission all values
must be multiplied by $\epsilon$. If the $\rm H_2$ emission is thermalized at $\rm 515~K$ and directly observed then 
the integrated line flux expected for 
the $\rm 2.12~\mu m$ (1--0) S(1) line is $\rm 7.5\times 10^{-13}~erg~cm^{-2}~s^{-1}$, which should be observable. If the $\rm H_2$ lines are scattered
into the line of sight with an efficiency of $\epsilon$, then no $\rm H_2$ lines should be visible in the $M$-band. Consequently, the $\rm H_2$ line
does not seem to be directly associated with the same gas emitting the CO lines since it is at least a factor of 10 too bright, but it may 
be an indication of a warmer component or the line may be
pumped. Sensitive observations of $\rm H_2$ lines in the K and L band
are needed to unambigously determine the origin of the molecular hydrogen emission.

\cite{Neufeld} show that if a dense gas of $10^{7.5}<n\rm_{H_2}<10^{12}~cm^{-3}$ 
is shocked, the vibrational emission from $\rm H_2$ is significantly suppressed partly because the shock dissociates the $\rm H_2$ molecules
and partly because the reforming molecules can easily be collisionally de-excited in the dense post-shock gas. The main coolants will be rotational 
and vibrational lines from $\rm H_2O$, $\rm OH$ and $\rm CO$. 
This is valid for a wide range of shock velocities $5<v_{\rm s}<100~\rm km~s^{-1}$ as long as the shock is of jump-type. 
A J-shock occurs in any case for $v_{\rm s}\gtrsim \rm 30~km~s^{-1}$ and for smaller shock velocities if the magnetic field is weak or if the 
length scale over which the shock occurs is small compared to the length scale for acceleration of charged particles.
Furthermore, the theoretical post-shock temperature structure
gives an interesting prediction. A J-shock will dissociate the gas at the shock front. The gas then cools slowly  from $\rm 10^5~K$ 
principally through the $\rm Ly\alpha$-line until $\rm CO$ forms at around $\rm 7000~K$ and $N\rm(HI)\sim 10^{21}~cm^{-2}$. Once the CO molecules are available
the gas cools quickly to about $\rm 500~K$ through the $\rm CO$ vibrational lines, where a constant temperature plateau is maintained due to the 
release of chemical potential energy as $\rm H_2$ is formed which is balanced against the thermalized vibrational emission from molecules.
The plateau occurs at an atomic hydrogen column density of $\rm 10^{21}-10^{22}~cm^{-2}$, which in the model by \cite{Neufeld} corresponds to a 
$\rm CO$ column density of $\rm 3\times 10^{17}-3\times 10^{18}~cm^{-2}$. It is stressed that the slower C-shock scenario was not treated
in such high density models, and it may produce similar cooling lines. Further modelling is needed to exclude a slow shock.
It is evident that the model values are remarkably similar to the values observed 
in GSS 30 IRS1 with temperature $T_{\rm gas}\rm =515~K$ and column density $N\rm (CO)=2\times 10^{18}~cm^{-2}$. However, it is unclear why no
hot $\rm 5000-7000~K$ component in the fundamental CO band is observed, which is also expected from the shock. Also 
no overtone emission is seen in the $K$-band spectrum of \cite{GL}. One plausible explanation is that the accretion activity is 
highly variable and that the hot component is only seen as the shock is occurring while the warm thermalized gas is seen in the post-shock medium. 
This would be supported by the variable water maser emission. The single observed $\rm H_2$ line is not inconsistent with an accretion shock although
it is too bright to be associated with a dense shocked gas and may only be an indication of a warmer less dense component. 

As described in \cite{Neufeld}, the most likely site for a dense shock in an embedded 
YSO is in a circumstellar disk which is accreting matter. When infalling matter is colliding with the disk at supersonic velocities, a J-shock is expected 
to occur at the surface of the disk. If the accretion velocities are too low to produce a J-shock, the molecules will not be dissociated and the observed
temperature plateau will not be present. The main uncertainty is if the high infall velocities required to create a dissociative shock are possible. However, infall velocities of $\rm 10~km~s^{-1}$ within $\rm 10~AU$ are reasonable \citep{Cassen}. 
Therefore, if the dissociative accretion shock 
scenario holds, the intrinsic line widths are {\it predicted} to be larger than or very close to $\rm 10~km~s^{-1}$. An accretion shock
at a large distance ($\rm \gtrsim 10~AU$) from the central star is consistent with the observation that the line emitting region is physically separated 
from the continuum emitting region. A shock will not heat a large amount of dust and the continuum emission, which has a colour temperature of
about $\rm 700~K$, is likely to be produced by irradiated dust within $\rm 1~AU$ of the star.

Other observable tracers of a shock in a dense medium are the rovibrational lines from the $\rm H_2O~\nu_2$ bending mode around $\rm 6~\mu m$ as well as
the vibrational transitions of $\rm OH$, all of which are unavailable from ground-based instruments. However, they may be detected
by the upcoming Stratospheric Observatory For Infrared Astronomy (SOFIA) mission.   

The possible dissociation of the molecules has important consequences for the chemistry in the inner disk. If the emission is indeed coming from an
accretion shock, and if a significant amount of matter passes through the shock, it will significantly affect the 
models of chemical evolution of the innermost parts of disks around low mass YSOs, by evaporating ice mantles and inducing high temperature chemistry.

 \section{Summary}

In this paper, we have analyzed the $\rm 4.5-4.8~\mu m$ spectrum of the embedded young stellar object GSS 30 IRS1 in the $\rho$ Ophiuchus core.
The observed emission lines from the rovibrational transitions of $\rm CO$ are fitted with a simple full radiative transfer 1D model. 
It is found that the line
emission must be scattered on a bipolar cavity in order to simultaneously account for the size
of the observed emitting region, the absolute flux level of the lines and the ratio of $\rm^{12}CO$ to $\rm^{13}CO$ lines. 
In this case the emission is well fitted by a single temperature gas with $T=\rm 515\pm5~K$ and a column density of 
$N\rm (CO)=2\pm0.5\times 10^{18}~cm^{-2}$. Furthermore, assuming a two-temperature distribution of the gas does not yield 
satisfactory fits. 

The observed emission line spectrum can be best explained by the presence of a J-shock in a dense medium, although other possible scenarios
cannot be ruled out. 
Most likely the shock is produced by the
accretion of gas onto a dense disk within a few tens of AU from the star as predicted by \cite{Neufeld}. Only this scenario explains why a large amount of
$\rm CO$ gas is thermalized at a single temperature of $\rm \sim 500~K$ and why the emission lines are so narrow. 

\begin{acknowledgements}
The authors wish to thank the VLT staff for all their help in obtaining the observations and in particular Chris Lidman for 
many helpful comments on the data reduction. We also acknowledge the many constructive suggestions made by an anonymous referee, which
helped improve the quality of this paper.
This research was supported by the Netherlands Organization for
Scientific Research (NWO) grant 614.041.004, the Netherlands Research 
School for Astronomy (NOVA) and a NWO Spinoza grant.
\end{acknowledgements}

\bibliographystyle{aa}
\bibliography{ms2585}
\end{document}